\documentclass[aps,prl,twocolumn,groupaddress,longbibliography,preprintnumbers]{revtex4-1} 

\usepackage{amsmath}
\usepackage{amsfonts}
\usepackage{amssymb}
\usepackage{graphicx}
\usepackage{color}
\usepackage{accents}
\usepackage[colorlinks=true, citecolor=cyan]{hyperref}

\newcommand{\Tr}{\mbox{Tr}}

\newcommand{\Sys}{\mathcal{S}}

\newcommand{\ket}[1]{|#1\rangle}
\newcommand{\bra}[1]{\langle #1|}

\makeatletter 
\newsavebox{\@brx}
\newcommand{\llangle}[1][]{\savebox{\@brx}{\(\m@th{#1\langle}\)}%
  \mathopen{\copy\@brx\kern-0.5\wd\@brx\usebox{\@brx}}}
\newcommand{\rrangle}[1][]{\savebox{\@brx}{\(\m@th{#1\rangle}\)}%
  \mathclose{\copy\@brx\kern-0.5\wd\@brx\usebox{\@brx}}}
\makeatother

\newlength{\dhatheight} 

\newcommand{\qed}{\nobreak \ifvmode \relax \else
      \ifdim\lastskip<1.5em \hskip-\lastskip
      \hskip1.5em plus0em minus0.5em \fi \nobreak
      \vrule height0.75em width0.5em depth0.25em\fi}

\begin{document}

\title{Comment on ``Thermodynamic principle for quantum metrology''}
\author{Shane Dooley}
\email[]{dooleysh@gmail.com}
\affiliation{School of Physics, Trinity College Dublin, College Green, Dublin 2, Ireland}
\author{Michael J. Kewming}
\affiliation{School of Physics, Trinity College Dublin, College Green, Dublin 2, Ireland}
\author{Mark T. Mitchison}
\affiliation{School of Physics, Trinity College Dublin, College Green, Dublin 2, Ireland}
\author{John Goold}
\affiliation{School of Physics, Trinity College Dublin, College Green, Dublin 2, Ireland}
\date{\today}



\maketitle 

\emph{Summary.} In \emph{Phys. Rev. Lett.} 128, 200501 (2022) \cite{Chu-22} the authors consider the thermodynamic cost of quantum metrology. One of the main results is their Eq. 3: \begin{equation} \mathcal{S} \geq \log(2) \| \hat{h}_\lambda \|^{-2} F_Q [\psi_\lambda] , \label{eq:inequality} \end{equation} which purports to relate the Shannon entropy $\Sys$ of an optimal measurement (i.e., in the basis of the symmetric logarithmic derivative) to the quantum Fisher information $F_Q$ of the pure state $\ket{\psi_\lambda}$. However, we show below that in the setting considered by the authors we have $\Sys = \log(2)$ and $\| \hat{h}_\lambda \|^{2} =  \max_{\psi_\lambda} F_Q[\psi_\lambda]$, so that Eq. 3 reduces to the trivial inequality $\max_{\psi_\lambda} F_Q[\psi_\lambda] \geq F_Q[\psi_\lambda]$, and does not in fact relate the entropy $\Sys$ to the quantum Fisher information. Moreover, for pure state quantum metrology, there exist optimal measurements (though not in the basis of the symmetric logarithmic derivative) for which $0 \leq \mathcal{S} \leq \log(2)$, leading to violations of Eq. \ref{eq:inequality} for some states $\ket{\psi_\lambda}$.

\emph{Background.} In deriving Eq. \ref{eq:inequality}, the authors restrict their analysis to pure states $\ket{\psi_\lambda} = \hat{U}_\lambda \ket{\psi}$ for some unitary $\hat{U}_\lambda$, where $\lambda$ is an unknown parameter to be estimated. If we measure $\ket{\psi_\lambda}$ with the POVM $\{ \hat{\Pi}_\alpha \}$, the estimation error is bounded by the Cram\'{e}r-Rao inequality $\delta\lambda \geq 1/\sqrt{F[\psi_\lambda, \{ \hat{\Pi}_\alpha \} ] }$, where: \begin{eqnarray} F [\psi_\lambda, \{ \hat{\Pi}_\alpha \} ] &=& \sum_\alpha \{ \partial_\lambda \langle \psi_\lambda | \hat{\Pi}_\alpha | \psi_\lambda \rangle \}^2 / \langle \psi_\lambda | \hat{\Pi}_\alpha | \psi_\lambda \rangle , \\ &=& \frac{1}{4} \sum_\alpha  \{ \Tr [\hat{L}_\lambda \hat{\Pi}_\alpha] \}^2 / \langle \psi_\lambda | \hat{\Pi}_\alpha | \psi_\lambda \rangle \label{eq:classical_FI} \end{eqnarray} is the (classical) Fisher information, and we have defined the (pure state) symmetric logarithmic derivative (SLD): \begin{equation} \hat{L}_\lambda = 2 \ket{\psi_\lambda}\bra{\partial_\lambda\psi_\lambda} + 2 \ket{\partial_\lambda\psi_\lambda}\bra{\psi_\lambda} . \label{eq:pure_SLD} \end{equation} It is known that the (classical) Fisher information can be maximised (with respect to the choice of measurement) by choosing $\{ \hat{\Pi}_\alpha \}$ to be the projectors onto the eigenbasis of $\hat{L}_\lambda$ \cite{Bra-94}. From Eq. \ref{eq:pure_SLD} it is clear that that the support of $\hat{L}_\lambda$ is a two-dimensional subspace that is spanned by the orthonormal states $\ket{\psi_\lambda}$ and $|\psi_\lambda^\perp \rangle = \mathcal{N} (\hat{\mathbb{I}} - \ket{\psi_\lambda} \bra{\psi_\lambda}) \ket{\partial_\lambda \psi_\lambda}$, where $\mathcal{N} = 1/\sqrt{\langle \partial_\lambda\psi_\lambda | \partial_\lambda\psi_\lambda \rangle - |\langle \partial_\lambda\psi_\lambda | \psi_\lambda \rangle |^2}$ ensures normalisation of $\ket{\psi_\lambda^\perp}$. In this basis the SLD is: \begin{equation} \hat{L}_\lambda = \frac{2}{\mathcal{N}} ( |\psi_\lambda \rangle \langle \psi_\lambda^\perp | + | \psi_\lambda^\perp \rangle \langle \psi_\lambda | ) , \end{equation} with the eigenstates $\ket{\pm} = (\ket{\psi_\lambda} \pm \ket{\psi_\lambda^\perp})/\sqrt{2}$ (all other eigenstates of $\hat{L}_\lambda$ have zero eigenvalue, and have zero overlap with $\ket{\psi_\lambda}$). The optimal POVM therefore includes the projectors $\hat{\Pi}_\pm = \ket{\pm}\bra{\pm}$, and for such a measurement we obtain the \emph{quantum} Fisher information (QFI): \begin{eqnarray} F_Q [\psi_\lambda] &=& \max_{\{ \hat{\Pi}_\alpha \}} F[\psi_\lambda, \{ \hat{\Pi}_\alpha \}] \\ &=& 4 \langle \partial_\lambda\psi_\lambda | \partial_\lambda\psi_\lambda \rangle - 4 |\langle \partial_\lambda\psi_\lambda | \psi_\lambda \rangle |^2 . \label{eq:QFI} \end{eqnarray} It was shown in Ref. \cite{Boi-07a} that a further optimization over the input state $\ket{\psi}$ gives the ultimate limit of pure-state quantum metrology, $\| \hat{h}_\lambda \|^{2} =  \max_{\psi} F_Q[\hat{U}_\lambda|\psi\rangle]$, where $\hat{h}_\lambda = i\hat{U}_\lambda^\dagger \partial_\lambda\hat{U}$ and $\| \bullet \|$ is the operator seminorm (the difference between the maximum and minumum eigenvalues of $\bullet$).

\emph{Entropy of optimal measurement.} The Shannon entropy corresponding to the outcome of the POVM $\{ \hat{\Pi}_\alpha \}$ is $\Sys [\{ \hat{\Pi}_\alpha \}] = - \sum_\alpha \langle \psi_\lambda | \hat{\Pi}_\alpha | \psi_\lambda \rangle \log \langle \psi_\lambda | \hat{\Pi}_\alpha | \psi_\lambda \rangle$. For an optimal measurement in the eigenbasis $\ket{\pm}$ of the SLD, we have the measurement outcome probabilities $p_\pm = |\langle\psi_\lambda | \pm \rangle|^2 = 1/2$. For pure-state quantum metrology, the Shannon entropy of such a measurement is therefore always $S = \log(2)$. This, combined with $\| \hat{h}_\lambda \|^{2} =  \max_{\psi} F_Q[\hat{U}_\lambda|\psi\rangle]$, shows that the inequality \ref{eq:inequality} reduces to the trivial inequality $\max_\psi F_Q[\psi_\lambda] \geq F_Q[\psi_\lambda]$.

\emph{Violation of the inequality.} For pure-state quantum metrology, the eigenbasis of $\hat{L}_\lambda$ is not the only measurement basis that maximises the (classical) Fisher information \cite{Bra-94}. Consider a POVM with the projectors $\hat{\Pi}_q = \ket{q}\bra{q}$ and $\hat{\Pi}_{\bar{q}} = \ket{\bar{q}}\bra{\bar{q}}$, where $\ket{q} = \sqrt{q} \ket{\psi_\lambda} + \sqrt{1-q} \ket{\psi_\lambda^\perp}$, $\ket{\bar{q}} = \sqrt{1-q} \ket{\psi_\lambda} - \sqrt{q} \ket{\psi_\lambda^\perp}$, $0 \leq q \leq 1$. Substituting this projective measurement into Eq. \ref{eq:classical_FI} gives $F[\psi_\lambda, \{ \hat{\Pi}_q, \hat{\Pi}_{\bar{q}} \}] = F_Q[\psi_\lambda]$, i.e., such a measurement maximises the (classical) Fisher information for any value of $q$. However, the Shannon entropy of this measurement is $\Sys = -q\log q - (1-q)\log(1-q)$, which can fall anywhere in the range $0 \leq \Sys \leq \log(2)$ depending on the value of $q$. This shows that the inequality \ref{eq:inequality} can be violated. We now present a simple example showing this violation.

\emph{Single-qubit example.} Consider the $\lambda$-dependent single-qubit state $\ket{\psi_\lambda} = e^{-i\lambda\hat{\sigma}_z/2} (\ket{0} + \ket{1})/\sqrt{2}$, where $\hat{\sigma}_z = \ket{0}\bra{0} - \ket{1}\bra{1}$. Using Eq. \ref{eq:QFI} the quantum Fisher information of this state is calculated to be $F_Q[\psi_\lambda] = 1$, and $\| \hat{h}_\lambda \|^2 = \| \hat{\sigma}_z / 2 \|^2 = 1$. Now consider a projective measurement of the observable $\hat{\sigma}(\phi) = \hat{\sigma}_x\cos\phi + \hat{\sigma}_y\sin\phi$ (i.e., the measurement projects onto its eigenstates $(\ket{0} \pm e^{i\phi}\ket{1})/\sqrt{2}$), which gives the outcome probabilities $p_\pm = \frac{1}{2}[1 \pm \cos(\phi - \lambda)]$ and the Shannon entropy $\Sys = -p_+ \log p_+ - p_- \log p_-$. Using Eq. \ref{eq:classical_FI} it is also straightforward to calculate the (classical) Fisher information of this measurement as $F = 1 = F_Q$, i.e., the measurement is optimal, independent of the measurement angle $\phi$. However, choosing $\phi = \lambda$, for example, gives: $\Sys = 0 \not\geq \log(2) = \log(2) \|\hat{h}_\lambda\|^2 F_Q [\psi_\lambda]$, clearly violating inequality \ref{eq:inequality}.






\bibliography{/Users/dooleysh/Google_Drive/physics/papers/bibtex_library/refs}

\begin{thebibliography}{3}%
\makeatletter
\providecommand \@ifxundefined [1]{%
 \@ifx{#1\undefined}
}%
\providecommand \@ifnum [1]{%
 \ifnum #1\expandafter \@firstoftwo
 \else \expandafter \@secondoftwo
 \fi
}%
\providecommand \@ifx [1]{%
 \ifx #1\expandafter \@firstoftwo
 \else \expandafter \@secondoftwo
 \fi
}%
\providecommand \natexlab [1]{#1}%
\providecommand \enquote  [1]{``#1''}%
\providecommand \bibnamefont  [1]{#1}%
\providecommand \bibfnamefont [1]{#1}%
\providecommand \citenamefont [1]{#1}%
\providecommand \href@noop [0]{\@secondoftwo}%
\providecommand \href [0]{\begingroup \@sanitize@url \@href}%
\providecommand \@href[1]{\@@startlink{#1}\@@href}%
\providecommand \@@href[1]{\endgroup#1\@@endlink}%
\providecommand \@sanitize@url [0]{\catcode `\\12\catcode `\$12\catcode
  `\&12\catcode `\#12\catcode `\^12\catcode `\_12\catcode `\%12\relax}%
\providecommand \@@startlink[1]{}%
\providecommand \@@endlink[0]{}%
\providecommand \url  [0]{\begingroup\@sanitize@url \@url }%
\providecommand \@url [1]{\endgroup\@href {#1}{\urlprefix }}%
\providecommand \urlprefix  [0]{URL }%
\providecommand \Eprint [0]{\href }%
\providecommand \doibase [0]{http://dx.doi.org/}%
\providecommand \selectlanguage [0]{\@gobble}%
\providecommand \bibinfo  [0]{\@secondoftwo}%
\providecommand \bibfield  [0]{\@secondoftwo}%
\providecommand \translation [1]{[#1]}%
\providecommand \BibitemOpen [0]{}%
\providecommand \bibitemStop [0]{}%
\providecommand \bibitemNoStop [0]{.\EOS\space}%
\providecommand \EOS [0]{\spacefactor3000\relax}%
\providecommand \BibitemShut  [1]{\csname bibitem#1\endcsname}%
\let\auto@bib@innerbib\@empty
\bibitem [{\citenamefont {Chu}\ and\ \citenamefont {Cai}(2022)}]{Chu-22}%
  \BibitemOpen
  \bibfield  {author} {\bibinfo {author} {\bibfnamefont {Yaoming}\ \bibnamefont
  {Chu}}\ and\ \bibinfo {author} {\bibfnamefont {Jianming}\ \bibnamefont
  {Cai}},\ }\bibfield  {title} {\enquote {\bibinfo {title} {Thermodynamic
  principle for quantum metrology},}\ }\href {\doibase
  10.1103/PhysRevLett.128.200501} {\bibfield  {journal} {\bibinfo  {journal}
  {Phys. Rev. Lett.}\ }\textbf {\bibinfo {volume} {128}},\ \bibinfo {pages}
  {200501} (\bibinfo {year} {2022})}\BibitemShut {NoStop}%
\bibitem [{\citenamefont {Braunstein}\ and\ \citenamefont
  {Caves}(1994)}]{Bra-94}%
  \BibitemOpen
  \bibfield  {author} {\bibinfo {author} {\bibfnamefont {Samuel~L.}\
  \bibnamefont {Braunstein}}\ and\ \bibinfo {author} {\bibfnamefont
  {Carlton~M.}\ \bibnamefont {Caves}},\ }\bibfield  {title} {\enquote {\bibinfo
  {title} {Statistical distance and the geometry of quantum states},}\ }\href
  {\doibase 10.1103/PhysRevLett.72.3439} {\bibfield  {journal} {\bibinfo
  {journal} {Phys. Rev. Lett.}\ }\textbf {\bibinfo {volume} {72}},\ \bibinfo
  {pages} {3439--3443} (\bibinfo {year} {1994})}\BibitemShut {NoStop}%
\bibitem [{\citenamefont {Boixo}\ \emph {et~al.}(2007)\citenamefont {Boixo},
  \citenamefont {Flammia}, \citenamefont {Caves},\ and\ \citenamefont
  {Geremia}}]{Boi-07a}%
  \BibitemOpen
  \bibfield  {author} {\bibinfo {author} {\bibfnamefont {Sergio}\ \bibnamefont
  {Boixo}}, \bibinfo {author} {\bibfnamefont {Steven~T}\ \bibnamefont
  {Flammia}}, \bibinfo {author} {\bibfnamefont {Carlton~M}\ \bibnamefont
  {Caves}}, \ and\ \bibinfo {author} {\bibfnamefont {John~M}\ \bibnamefont
  {Geremia}},\ }\bibfield  {title} {\enquote {\bibinfo {title} {Generalized
  limits for single-parameter quantum estimation},}\ }\href@noop {} {\bibfield
  {journal} {\bibinfo  {journal} {Phys. Rev. Lett.}\ }\textbf {\bibinfo
  {volume} {98}},\ \bibinfo {pages} {090401} (\bibinfo {year}
  {2007})}\BibitemShut {NoStop}%
\end{thebibliography}%

\end{document}